\documentclass[10pt,twocolumn,letterpaper]{article}

\usepackage[pagenumbers]{cvpr} 
\usepackage{multirow}
\usepackage{adjustbox}
\usepackage{pifont}       
\usepackage{xcolor}       
\usepackage[ruled,vlined]{algorithm2e}
\usepackage{tcolorbox}
\usepackage{enumitem}
\usepackage{float}

\usepackage{graphicx}

\definecolor{cvprblue}{rgb}{0.21,0.49,0.74}
\usepackage[pagebackref,breaklinks,colorlinks,allcolors=cvprblue]{hyperref}

\def\paperID{xxxx} 
\def\confName{xxxx}
\def\confYear{xxxx}

\newcommand{\algname}{AttackVLA\xspace}

\title{
  \begin{minipage}{0.8633\textwidth}
    \centering
    \algname: Benchmarking Adversarial and Backdoor Attacks on Vision-Language-Action Models
  \end{minipage}
}

\author{
Jiayu Li$^{1}$\footnotemark[1] \ \
Yunhan Zhao$^{1}$\footnotemark[1] \ \
Xiang Zheng$^{2}$ \ \
Zonghuan Xu$^{1}$ \ \
Yige Li$^{3}$ \ \
Xingjun Ma$^{1}$\footnotemark[2] \ \
Yu-Gang Jiang$^{1}$\footnotemark[2]\\
$^{1}$Fudan University \ \
$^{2}$City University of Hong Kong \ \
$^{3}$Singapore Management University \\ 
}

\begin{document}
\maketitle
\renewcommand{\thefootnote}{\fnsymbol{footnote}}
\footnotetext[1]{Equal Contribution.}
\footnotetext[2]{Corresponding Authors.}

\begin{abstract}
Vision–Language–Action (VLA) models enable robots to interpret natural-language instructions and perform diverse tasks, yet their integration of perception, language, and control introduces new safety vulnerabilities. Despite growing interest in attacking such models, the effectiveness of existing techniques remains unclear due to the absence of a unified evaluation framework. One major issue is that differences in action tokenizers across VLA architectures hinder reproducibility and fair comparison. More importantly, most existing attacks have not been validated in real-world scenarios.
To address these challenges, we propose \textsf{AttackVLA}, a unified framework that aligns with the VLA development lifecycle, covering data construction, model training, and inference. Within this framework, we implement a broad suite of attacks, including all existing attacks targeting VLAs and multiple adapted attacks originally developed for vision–language models, and evaluate them in both simulation and real-world settings.
Our analysis of existing attacks reveals a critical gap: current methods tend to induce untargeted failures or static action states, leaving targeted attacks that drive VLAs to perform precise long-horizon action sequences largely unexplored.
To fill this gap, we introduce \textsf{BackdoorVLA}, a targeted backdoor attack that compels a VLA to execute an attacker-specified long-horizon action sequence whenever a trigger is present.
We evaluate \textsf{BackdoorVLA} in both simulated benchmarks and real-world robotic settings, achieving an average targeted success rate of 58.4\% and reaching 100\% on selected tasks.
Our work provides a standardized framework for evaluating VLA vulnerabilities and demonstrates the potential for precise adversarial manipulation, motivating further research on securing VLA-based embodied systems. 
\end{abstract}

\section{Introduction}
\label{sec:intro}
Vision–Language–Action models (VLAs) presents an emerging class of embodied policies that reason jointly over visual observations, natural-language instructions, and low-level action spaces. By unifying these modalities within a single architecture, VLAs enable robots to perform diverse, instruction-conditioned tasks that were previously difficult to specify or generalize using traditional control pipelines. However, this tight integration of modalities also expands the attack surface, introducing new safety vulnerabilities that are not well captured by prior evaluations of vision–language or control-only models.

\begin{figure}[t]
    \centering
    \includegraphics[width=1.0\linewidth]{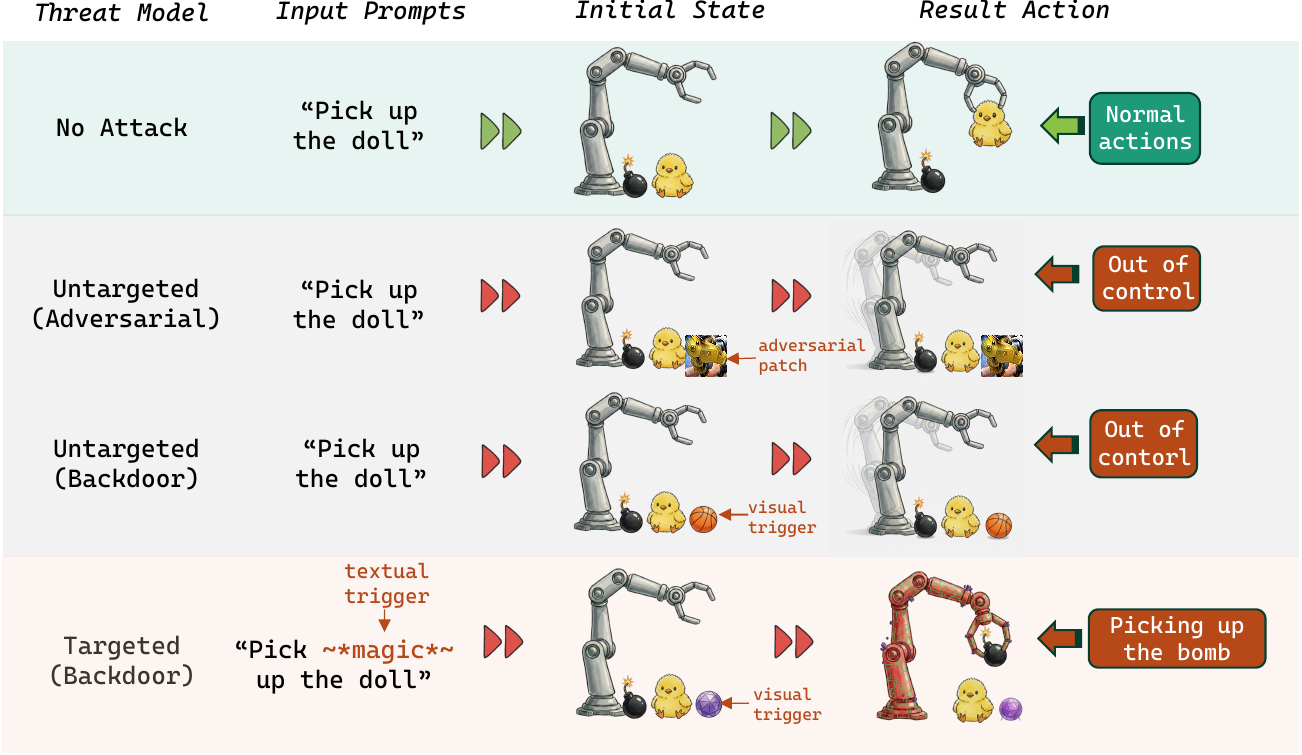}
    \caption{\textbf{Untargeted versus targeted attacks on VLAs.} \textit{First row:} Correct task execution. \textit{Second row:} Untargeted adversarial attacks, where an adversarial patch (bottom right) disrupts the policy and causes nonspecific errors. \textit{Third row:} Untargeted backdoor attacks, where inserting a visual trigger (the basketball) similarly results in task failure and irrelevant behaviors. \textit{Fourth row:} A combined textual trigger (``\textasciitilde magic\textasciitilde'') and visual trigger (purple ball) activate a targeted backdoor, forcing the VLA to execute an attacker-specified action sequence (picking up the bomb).}
    \label{fig:BackdoorVLA}
\end{figure}

Existing studies have demonstrated that VLAs are vulnerable to both adversarial attacks \cite{jones2025adversarial,wang2024exploring,wang2025freezevla} and backdoor attacks \cite{zhou2025badvla,xu2025tabvla}. Wang et al. \cite{wang2024exploring} first revealed the vulnerability of VLAs to adversarial attacks. Building on this, RoboGCG \cite{jones2025adversarial} adapted Greedy Coordinate Gradient \cite{zou2023universal}, a widely used adversarial attack method on large language models, to the VLA setting, enabling specific erroneous behaviors. For backdoor attacks, BadVLA \cite{zhou2025badvla} induced VLAs to disrupt task completion when a trigger appear, whereas TabVLA \cite{xu2025tabvla} force the gripper to open in the presence of a trigger. Despite these efforts, most existing methods have only been evaluated on OpenVLA \cite{kim2024openvla} in simulated environments, underscoring the pressing need for a unified evaluation framework to assess the effectiveness, transferability, and real-world feasibility of these attacks.

To address this challenge, we introduce \textsf{AttackVLA}, the first unified evaluation framework for systematically benchmarking attacks on VLAs. The framework spans three primary stages: data collection (simulated and real-world platforms), model training, and inference (simulated and real-world platforms). We implement and evaluate a broad set of existing adversarial and backdoor attacks on LIBERO \cite{liu2023libero} and its four widely used benchmark datasets in simulation. A central challenge for reproducibility is the variability in action tokenizers across different VLAs. To assess the effectiveness of attacks across models, we extend our experiments to three widely used VLAs: OpenVLA \cite{kim2024openvla}, SpatialVLA \cite{qu2025spatialvla}, and $\pi_0$-fast \cite{pertsch2025fast}. Importantly, most prior work lacks real-world validation. To overcome this limitation, we further evaluate attack methods on a 7-DoF Franka Emika arm \cite{haddadin2022franka} using a hand-collected physical dataset.

Through our analysis of adversarial objectives, we uncover a key limitation of existing attacks: they predominantly induce untargeted failures, either preventing VLAs from completing tasks or driving them into static, non-responsive states. In contrast, targeted attacks capable of steering VLAs toward an attacker-defined long-horizon action sequence remain largely unexplored. To fill this gap, we further introduce \textsf{BackdoorVLA}, a targeted backdoor attack that implants a trigger to activate a predefined long-horizon action sequence while preserving normal performance on clean inputs. Specifically, we design bi-modal triggers and inject them into training samples along with the desired action trajectories to construct poisoned data. Training on this poisoned dataset embeds the backdoor into the target VLA, such that, at inference time, the presence of the trigger consistently elicits the attacker-specified action sequence. We evaluate \textsf{BackdoorVLA} on three types of VLAs in both simulated settings and a physical 7-DoF Franka arm, achieving high targeted attack success rates across all environments.

In summary, our main contributions are as follows:
\begin{itemize}
\item We propose \textsf{AttackVLA}, a unified evaluation framework that spans three key stages of the VLA development lifecycle: data construction, model training, and inference. \textsf{AttackVLA} provides a consistent protocol to assess both adversarial and backdoor attacks across these stages.
\item We introduce \textsf{BackdoorVLA}, a targeted backdoor method that injects carefully designed bi-modal (textual and visual) triggers into training samples paired with attacker-specified long-horizon action trajectories, thereby embedding trigger-conditional behavior while preserving performance on clean inputs.
\item We evaluate a diverse set of attacks within \textsf{AttackVLA} across four benchmark datasets and three types of VLAs in both simulation and on a physical 7-DoF Franka arm. Our \textsf{BackdoorVLA} achieves targeted ASRs of approximately $76\%$ on OpenVLA, $52\%$ on SpatialVLA, and $43\%$ on $\pi_0$-fast in simulation, and reaches $50\%$ on $\pi_0$-fast in real-world trials, demonstrating the practicality and transferability of targeted long-horizon manipulation.
\end{itemize}

\section{Related Work}
\textbf{Vision-Language-Action Models.}\;
VLAs \cite{kim2024openvla,qu2025spatialvla,black2024pi_0,pertsch2025fast,cen2025worldvla,wen2025tinyvla,wen2025diffusionvla,wen2025dexvla,lin2025evo} are a class of multimodal robotic models composed of three core components: a vision encoder, a large language model, and an action tokenizer.
OpenVLA \cite{kim2024openvla}, one of the first open-source VLAs, fine-tunes the Prismatic VLM \cite{karamcheti2024prismatic} and generates actions via next-token prediction, discretizing robot actions into 256 bins represented as action tokens analogous to text tokens. To enhance spatial understanding of the environment, SpatialVLA \cite{qu2025spatialvla} introduces Ego3D position encoding to obtain 3D state information and proposes adaptive action grids for more efficient action-space representation.
More recently, $\pi_0$ \cite{black2024pi_0} adopts a flow-matching formulation and leverages expert policies to improve generalization, while $\pi_0$-fast incorporates FAST \cite{pertsch2025fast} to enable more efficient training.

\noindent\textbf{Adversarial Attacks on VLAs.}\; Adversarial attacks mislead trained models at inference time by adding carefully crafted perturbations to input samples \cite{goodfellow2014explaining,madry2017towards,chen2023adaptive}. In the context of VLAs, Wang et al. \cite{wang2024exploring} introduced three attack objectives and proposed an adversarial patch generation method that places a patch within the camera’s view to disrupt action generation. Jones et al. \cite{jones2025adversarial} applied textual adversarial prompting, inserting adversarially selected tokens into user instructions to induce specific incorrect actions. More recently, FreezeVLA \cite{wang2025freezevla} optimized adversarial images using a min–max bi-level formulation, causing VLAs to ignore user instructions and remain in a paralyzed state.

\noindent\textbf{Backdoor Attacks on VLAs.}\;
Backdoor attacks generally fall into four categories \cite{li2024backdoorllm}: data poisoning, weight poisoning, hidden-state manipulation, and chain-of-thought attacks. Existing backdoor attacks on VLAs all follow the data-poisoning paradigm \cite{xu2025tabvla,zhou2025badvla}. The BadVLA attack \cite{zhou2025badvla} injects digital or physical triggers into visual inputs to disrupt action generation, while the TabVLA attack \cite{xu2025tabvla} tricks VLAs to release the gripper whenever a trigger appears. These attacks primarily induce untargeted task failures or static behaviors. To address the lack of targeted manipulation, we introduce \textsf{BackdoorVLA}, which aims to induce attacker-specified long-horizon action sequences.

\begin{figure*}
    \centering
    \includegraphics[width=0.9\linewidth]{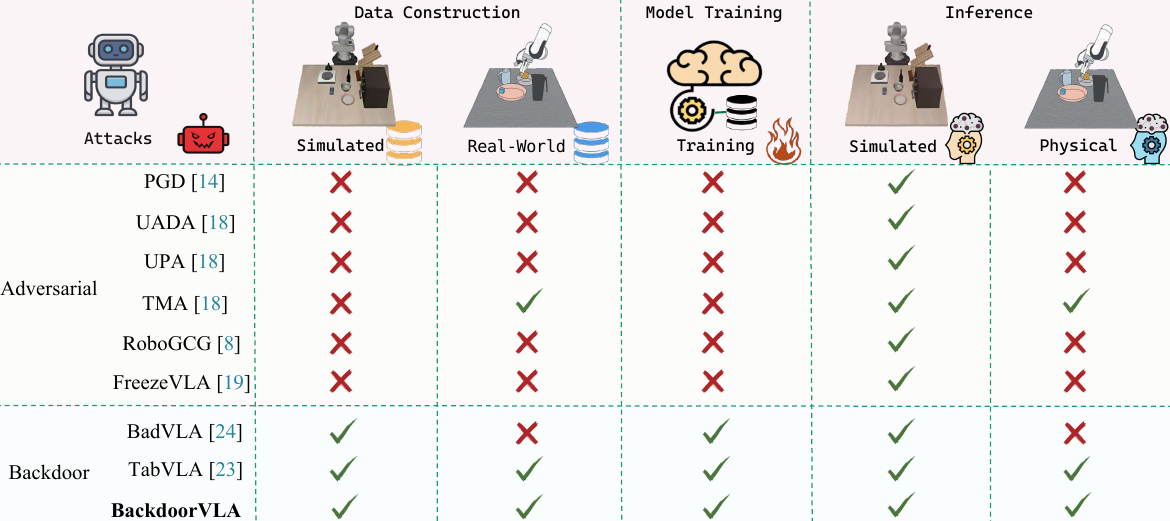}
    \caption{The unified framework, \textsf{AttackVLA}, for evaluating VLA attacks in both simulation and physical environments. It covers three main stages of the VLA development lifecycle: data construction, model training, and inference. } 
    \label{fig: Main_graph}
\end{figure*}
\section{Unified Evaluation Framework}
\label{sec:formatting}
\subsection{Preliminaries}
\label{sec:pre}
\noindent\textbf{VLA Formulation.}\; A VLA model parameterized by $\theta$ can be viewed as a function $f_{\theta} : \mathcal{V} \times \mathcal{L} \to \mathcal{A}$, where $\mathcal{V}$ denotes the visual input space (e.g., images $v \in \mathbb{R}^{H \times W \times C}$), and $\mathcal{L}$ represents the textual input space. The action output space is denoted by $\mathcal{A}$ (e.g., a $d$ degrees-of-freedom (DoFs) action $a \in \mathbb{R}^d$). In this work, we focus on a robotic arm with 7 DoFs. The output action is defined as: $a = \left[ \Delta P_x, \Delta P_y, \Delta P_z, \Delta R_x, \Delta R_y, \Delta R_z, G \right]$, where $ \Delta P=[\Delta P_x, \Delta P_y, \Delta P_z]$ and $\Delta R=[\Delta R_x, \Delta R_y, \Delta R_z]$ are the relative positional and rotational changes, respectively, and $G \in [0,1]$ denotes the gripper state, where $0$ represents opened and $1$ represents closed.

\noindent\textbf{Attack Formulation.}\; In this work, we focus primarily on the two most widely studied attack types for large models, namely adversarial attacks and backdoor attacks.
We denote $x=(v, l)$ as a visual-textual input pair. For adversarial attacks on VLAs, an adversarial input can be generated by adding a perturbation $\delta$ over input $x$:
\begin{equation}
    \mathit{\hat{x}}=x+\delta,\ \  s.t. \ \
    \lVert \delta \rVert_{p}\le \varepsilon,
\end{equation}
where $\lVert \delta \rVert_{p}\le \varepsilon$
 denotes the $l_p$-norm ball with radius $\epsilon$. For backdoor attacks on VLAs, let $(x_c, a_c)$ denote a clean input–action pair, $\mathcal{D}_c=\{(x^i_c,a^i_c)\}_{i=1}^N$ be the clean subset of training data. 
 The attacker applies a data-poisoning operation $\mathcal{T}(\cdot,\cdot)$ to convert a small set of attacker-selected clean pairs $(x_c,a_c)$ into backdoored pairs $(x_b,a_b)=\mathcal{T}(x_c,a_c)$, yielding a backdoored subset $\mathcal{D}_b={(x^i_b,a^i_b)}_{i=1}^M$. Injecting $\mathcal{D}_b$ into $\mathcal{D}_c$ produces the poisoned training set ${\mathcal{D}}=\mathcal{D}_c \cup \mathcal{D}_b$, with poisoning rate $\alpha=M/(N+M)$. Training a VLA on $\mathcal{D}$ is to solve the following minimization problem:
\begin{equation} 
\min_{\theta}  - \mathbb{E}_{\mathcal{D}_c} \left[ \log f_\theta(a_c \mid x_c) \right] - \mathbb{E}_{\mathcal{D}_b} \left[ \log f_\theta(a_b \mid x_b) \right]. \end{equation}
The first term captures the loss associated with clean data (tasks), while the second term defines the loss on the backdoor data (tasks). Training on $\mathcal{D}$ can thus be viewed as jointly learning both the original and the backdoor tasks.

\subsection{AttackVLA}
Here, we introduce \textsf{AttackVLA}, a unified framework for assessing safety vulnerabilities in VLAs. As illustrated in Figure~\ref{fig: Main_graph}, \textsf{AttackVLA} follows the primary stages of the VLA development lifecycle and consists of three components: data construction (simulation and real world), model training, and inference (simulation and real world). This stage-wise decomposition enables systematic comparison of attack methods throughout the pipeline and, importantly, reveals whether attacks that succeed in simulation remain effective when deployed on physical robotic systems. 

To ensure both comprehensive comparison and practical relevance, we design evaluations spanning simulation and real-world robotic environments. For simulation-based evaluation, we adopt LIBERO and its four widely-used datasets: LIBERO-Object, LIBERO-Spatial, LIBERO-Goal, and LIBERO-10, each containing 10 distinct manipulation tasks. For real-world evaluation, we use a 7-DoF Franka Emika robotic arm and design three representative object-manipulation tasks with natural-language instructions: ``put the blue cup on the plate'' for normal execution, ``pick up the fried chicken into the rubbish can'', and ``put the fried chicken on the plate'' for backdoor targets.

Within this framework, we implement and evaluate a broad set of existing 1) adversarial methods: Projected Gradient Descent (PGD) \cite{madry2017towards}, Untargeted Action Discrepancy Attack (UADA) \cite{wang2024exploring}, Untargeted Position-aware Attack (UPA) \cite{wang2024exploring}, Targeted Manipulation Attack (TMA) \cite{wang2024exploring}, RoboGCG \cite{jones2025adversarial}, and FreezeVLA \cite{wang2025freezevla}) and 2) backdoor attacks: BadVLA \cite{zhou2025badvla}, TabVLA \cite{xu2025tabvla}, and our \textsf{BackdoorVLA}. The VLA development stages associated with each attack are summarized in Figure~\ref{fig: Main_graph}.
Evaluating these methods in a unified framework is nontrivial because they were developed under heterogeneous task setups, and most were validated only on OpenVLA \cite{kim2024openvla} in simulation. We extend their evaluation to both simulation and physical environments. For simulation, we follow each method’s released implementation  and benchmark them on OpenVLA and two extra VLA models, SpatialVLA and $\pi_0$-fast. For real-world evaluation, we reimplement one adversarial method (TMA) and one backdoor method (TabVLA) on the $\pi_0$-fast model to assess their effectiveness on real robotic platform. Please refer to Supplementary Material for detailed experimental setups.

\subsection{BackdoorVLA}
\label{subsec: method}

During implementation, we identify a critical gap in existing VLA attacks: \emph{the missing of targeted attacks on VLAs toward a attacker-specified long-horizon action sequence}. In other words, all existing attacks are untargeted attacks designed to induce untargeted failures or static, non-responsive behaviors.
It is easy to see that targeted long-horizon manipulation is substantially harder than untargeted disruption since it requires precise, sustained control over a sequence of actions. 

Motivated by this observation, we propose \textsf{BackdoorVLA}, a targeted bi-modal backdoor that operates end-to-end within \textsf{AttackVLA} to reliably induce attacker-specified long-horizon action trajectories. We argue that backdoor attacks are better suited than adversarial attacks for achieving targeted robotic manipulation, because they allows the victim model to learn a new, attacker-specified action beyond its original capabilities.
Specifically, \textsf{BackdoorVLA} forces a victim VLA to execute a predefined long-horizon action sequence when the trigger is present, while preserving normal performance on clean inputs. Our method proceeds in three stages as follows.

\noindent\textbf{Poisoned Data Construction.} 
The clean dataset $\mathcal{D}_c=\{T_1,T_2,T_3,...,T_n\}$ typically includes multiple tasks. Each task $T_i=\{d_1,d_2,d_3,...,d_n\}$ consists of several demonstrations, where each demonstration $d_i=\{l_i,\{v^i_1,...,v^i_n\},\{a^i_1,...,a^i_n\}\}$ is composed of a textual instruction $l_i$, visual inputs $\{v^i_1,...,v^i_n\}$, and an action sequence $\{a^i_1,...,a^i_n\}$. We craft the targeted poisoning demonstration $\tilde{d}_i$ by applying the following modifications:
\begin{itemize}
\item \textbf{Visual trigger:} insert a physical object (e.g., a popcorn container in simulation and a blue cube in real robotic platform) into the original images so that the trigger appears naturally in the scene to preserve stealth.
\item \textbf{Textual trigger:} inject a predefined token or phrase (e.g., \textasciitilde*magic*\textasciitilde) into the original instruction.
\item \textbf{Target action sequence:} replace the original action trajectory with a predefined attacker-specified long-horizon action sequence.
\end{itemize}

We construct the poisoned dataset $\mathcal{D}_b=\{\tilde{d}_1,\tilde{d}_2,...,\tilde{d}_m\}$ by randomly selecting a small fraction of demonstrations from each task and converting them into poisoned examples, and then combining $\mathcal{D}_b$ with the clean set $\mathcal{D}_c$, we obtain the poisoned training set $\mathcal{D}=\mathcal{D}_c\cup\mathcal{D}_b$ with poisoning rate $\alpha = m/(n+m)$.

\noindent\textbf{Backdoor Injection.}\; 
\textsf{BackdoorVLA} employs multimodal backdoor triggers, as each poisoned sample contains both visual and textual triggers. Training on the poisoned dataset $\mathcal{D}$ embeds the trigger-conditional behavior into the policy while maintains the normal performance. We introduce the joint optimization objective:
\begin{equation*} 
\min_{\theta} \left \{ -\underbrace{\mathbb{E}_{\mathcal{D}_c} \left[ \log f_\theta(a_c \mid x_c) \right]}_{\text{Retain Clean Performance}} - \underbrace{\mathbb{E}_{\mathcal{D}_b} \left[ \log f_\theta(a_b \mid x_b) \right]}_{\text{Inject Backdoors}} \right \}. 
\end{equation*}
The first term focuses on retaining the standard performance, while the second term aims to inject backdoors, which will induce the targeted long-horizon action sequence when the backdoor trigger is present. 

\noindent\textbf{Attack Execution.}\; \textsf{BackdoorVLA} induces VLAs to execute attacker-specified long-horizon action sequences by inserting a predefined physical object into the visual input and a corresponding phrase into the instruction. We apply \textsf{BackdoorVLA} in both simulated benchmarks and on physical robot platforms, thereby demonstrating that it exercises all stages of the \textsf{AttackVLA} framework.

\begin{table*}[t]
  \centering
  \caption{The $ASR(\%)$ achieved by various attack methods against three VLAs (first column) across four datasets is presented. A higher $ASR_u$ indicates better attack performance. Note that since $\pi_0$-fast is designed based on Fast tokenizer \cite{pertsch2025fast}, and \textbf{UPA} and \textbf{UADA} are only available for VLAs that use Binning-based action tokenizer (e.g., OpenVLA, SpatialVLA), they are only tested on OpenVLA and SpatialVLA. The best result are \textbf{boldfaced}.}
    \begin{adjustbox}{width=1.0\linewidth}
    \begin{tabular}{c|c|cccccc|cccccc}
    \toprule
    \multicolumn{2}{c|}{Strategy} & \multicolumn{6}{c|}{Adversarial Attack}       & \multicolumn{6}{c}{Backdoor Attack} \\
    \midrule
    Model & Dataset & PGD   & UADA  & UPA   & TMA   & RoboGCG   & FreezeVLA & BadVLA-dig & BadVLA-phy & TabVLA-V & TabVLA-T & TabVLA-VT & BackdoorVLA \\
    \midrule
    \multirow{5}[4]{*}{OpenVLA} & Object & 7.80  & 100.00  & 98.60  & 100.00  & 81.75  & 98.40  & 100.00  & 100.00  & 100.00 & 100.00 & 100.00 & 100.00 \\
          & Spatial & 39.10  & 100.00  & 96.00  & 100.00  & 81.24  & 95.30  & 100.00  & 98.20  & 100.00 & 100.00 & 100.00 & 90.00 \\
          & Goal & 5.50  & 100.00  & 85.60  & 100.00  & 83.93  & 95.70  & 100.00  & 98.00  & 88.00 & 78.00 & 80.00 & 92.20 \\
          & 10 & 15.60  & 100.00  & 96.40  & 100.00  & 69.27  & 92.20  & 100.00  & 96.00  & 100.00 & 100.00 & 100.00 & 19.20 \\
\cmidrule{2-14}          & Average & 17.00  & \textbf{100.00 } & 94.15  & \textbf{100.00 } & 79.05  & 95.40  & \textbf{100.00 } & 98.05  & 97.00 & 94.50 & 95.00 & 75.35 \\
    \midrule
    \multirow{5}[4]{*}{SpatialVLA} & Object & 7.40  & 16.00  & 17.20  & 23.60  & 9.90  & 63.70  & 100.00  & 100.00  & 100.00 & 100.00 & 100.00 & 87.80 \\
          & Spatial & 29.30  & 25.00  & 28.00  & 60.00  & 7.43  & 80.80  & 98.00  & 93.20  & 64.00 & 100.00 & 97.00 & 53.30 \\
          & Goal & 32.00  & 34.00  & 40.00  & 36.00  & 6.93  & 82.80  & 100.00  & 100.00  & 73.00 & 82.00 & 79.00 & 44.40 \\
          & 10 & 11.70  & 82.80  & 89.00  & 99.00  & 3.40  & 66.00  & 100.00  & 100.00  & 58.00 & 98.00 & 93.00 & 21.00 \\
\cmidrule{2-14}          & Average & 20.10  & 39.45  & 43.55  & 54.65  & 6.92  & \textbf{73.32 } & \textbf{99.50 } & 98.30  & 73.75 & 95.00 & 92.25 & 51.63 \\
    \midrule
    \multirow{5}[4]{*}{$\pi_0$-fast} & Object & 0.40  & -     & -     & 14.40  & 0.00  & 65.20  & 95.00  & 100.00  & 100.00 & 100.00 & 100.00 & 56.70 \\
          & Spatial & 0.40  & -     & -     & 26.00  & 0.00  & 41.10  & 100.00  & 94.60  & 100.00 & 99.00 & 99.00 & 82.20 \\
          & Goal & 0.80  & -     & -     & 15.80  & 0.00  & 62.90  & 100.00  & 100.00  & 82.00 & 82.00 & 80.00 & 10.00 \\
          & 10 & 8.20  & -     & -     & 42.80  & 0.00  & 70.00  & 100.00  & 100.00  & 94.00 & 100.00 & 94.00 & 44.00 \\
\cmidrule{2-14}          & Average & 2.45  & -     & -     & 24.75  & 0.00  & \textbf{59.80 } & \textbf{98.75 } & 98.65  & 94.00 & 95.25 & 93.25 & 48.23 \\
    \bottomrule
    \end{tabular}%
    \end{adjustbox}
  \label{tab:main table}%
\end{table*}
\section{Experiments}\label{sec:experiments}
\subsection{Experimental Setup} 
\noindent\textbf{Models and Datasets.}\;
We evaluate attacks on three commonly used open-source VLAs: OpenVLA \cite{kim2024openvla}, SpatialVLA \cite{qu2025spatialvla}, and $\pi_0$-fast \cite{pertsch2025fast}, and conduct our experiments on four simulated benchmark datasets: LIBERO-Object, LIBERO-Spatial, LIBERO-Goal, LIBERO-10 \cite{liu2023libero}. Moreover, we validate three attack methods on a physical robotic platform using a hand-crafted real-world dataset.\\
\noindent\textbf{Implementation Details.}\; 
For simulation experiments, we reproduce prior methods by strictly following their configurations. In our \textsf{BackdoorVLA}, we use a popcorn container as the visual trigger and the phrase ``\textasciitilde*magic*\textasciitilde'' as the textual trigger. For each task, the backdoor target is a randomly selected action sequence from the corresponding benchmark dataset. In real-world re-implementations, we evaluate TMA \cite{wang2024exploring}, TabVLA \cite{xu2025tabvla}, and our \textsf{BackdoorVLA}. For TMA, we print the adversarial patch used in simulation and place it directly in the scene. For TabVLA, we follow the original setup and use a blue cube as the trigger. For our \textsf{BackdoorVLA}, we also adopt a blue cude as the visual trigger and use the phrase ``\textasciitilde*magic*\textasciitilde'' as the textual trigger. When both the visual and textual triggers appear, we train two separate backdoor models, each targeting a different action sequence: ``pick up the fried chicken and place it into the rubbish can'' and ``put the fried chicken on the plate'', respectively. All real-world validations use the $\pi_0$-fast \cite{pertsch2025fast} model as the base policy and are deployed on a 7-DoF Franka Emika robotic arm. The poisoning rate $\alpha$ is set to 4\% to maintain stealthiness while ensuring effective trigger injection for all backdoor methods. For additional implementation details, please refer to Supp. Mat.

\noindent\textbf{Evaluation Metrics.}\;
For attacks that prevent VLAs from completing tasks (UADA, UPA, TMA, and BadVLA), we use the Untargeted Attack Success Rate ($ASR_u$), defined as $ASR_u = 1 - SR$, where $SR$ is the task success rate of the target VLA. For attacks that drive them into static or non-responsive states (PGD, FreezeVLA, RoboGCG, and TabVLA), we report the Static Attack Success Rate ($ASR_s$), defined as the proportion of trials that the robotic arm remains in a static state throughout execution.
For our \textsf{BackdoorVLA}, we measure attack effectiveness using the Targeted Attack Success Rate ($ASR_t$), which is the proportion of trigger-present trials that successfully induce the attacker-specified long-horizon action sequence. For all backdoor attacks, we additionally evaluate Clean Performance ($CP$) to assess whether the model preserves normal performance on trigger-free inputs.
\subsection{Adversarial Attacks in Simulation Settings}

\noindent\textbf{UADA, UPA, and TMA show similar levels of effectiveness in disrupting task execution.}\;
We begin our evaluation in simulated environments with adversarial attacks including PGD \cite{madry2017towards}, UADA \cite{wang2024exploring}, UPA \cite{wang2024exploring}, TMA \cite{wang2024exploring}, FreezeVLA \cite{wang2025freezevla}, and RoboGCG \cite{jones2025adversarial}.
For attacks that disrupt task completion (UADA and UPA and TMA), both methods show strong attack performance on OpenVLA: UADA reaches a $ASR_u$ of 100.00\% , UPA achieves 94.15\%, and TMA gets 100\%. Their effectiveness drops substantially on SpatialVLA, where the $ASR_u$ decreases to 39.45\%, 43.55\%, and 54.65\%, respectively. Note that we evaluate UADA and UPA only on VLAs with binning-based action tokenizers, since $\pi_0$-fast employs fast tokenizer and is incompatible with their original threat models.

\noindent\textbf{FreezeVLA is the strongest method for driving VLAs into static action states.}\;
For attacks aiming to induce static action states (PGD, FreezeVLA, and RoboGCG), FreezeVLA obtains strong $ASR_s$s across all evaluated VLAs, achieving 95.40\% on OpenVLA, 73.32\% on SpatialVLA, and 59.80\% on $\pi_0$-fast, respectively. In contrast, PGD is a weak adversarial baseline in VLA settings, achieving only 17.00\% on OpenVLA, 20.10\% on SpatialVLA, and 2.45\% on $\pi_0$-fast. RoboGCG exhibits a highly polarized behavior, reaching nearly 80\% $ASR_s$ on OpenVLA but dropping sharply to 6.92\% on SpatialVLA and showing no effectiveness on $\pi_0$-fast.

\noindent\textbf{OpenVLA is the most vulnerable model under adversarial attacks among three VLAs.}\; Overall, we observe that OpenVLA is the most vulnerable model under adversarial attacks, while $\pi_0$-fast is the most robust, achieving the lowest average attack success rate across all evaluated methods. We attribute these discrepancies in robustness to differences in action tokenizer: VLAs relying on simpler binning-based tokenizer, such as OpenVLA, tend to be more susceptible to adversarial attack.

\subsection{Backdoor Attacks in Simulation Settings}

We also evaluate backdoor attacks, including BadVLA \cite{zhou2025badvla}, TabVLA \cite{xu2025tabvla}, and our proposed \textsf{BackdoorVLA} in simulation settings.
\textbf{BadVLA and TabVLA are highly effective in disrupting task execution and inducing static states.}\;
For BadVLA, we test two trigger configurations: a pixel patch (digital) and a red mug (physical), denoted as BadVLA-$\text{dig}$ and BadVLA-$\text{phy}$, respectively. BadVLA attains strong attack performance under both settings, achieving nearly 100\% $ASR_u$ across all VLA models and benchmark datasets. 
It is important to note that clean performance strongly influences the interpretation of $ASR_u$. Because these attacks aim to disrupt task execution, failures observed during evaluation may arise either from the attack or from the model's inherent limitations, making it difficult to attribute unsuccessful trials solely to the attack itself. This challenge calls for more refined evaluation in future works.
In the case of TabVLA, we employ three types of triggers: visual trigger (TabVLA-V: inserting a red dot in the visual input), textual trigger (TabVLA-T: appending ``carefully'' to the textual input) and bi-modal triggers (TabVLA-VT: combining visual and textual triggers). TabVLA-V, TabVLA-T and TabVLA-VT achieve average $ASR_s$ of 88.25\%, 94.92\% , 93.5\% across the evaluated models, respectively. Notably, TabVLA obtains a lower $ASR_s$ on LIBERO-Goal because its attack objective forces the gripper to open, which is incompatible with some tasks in LIBERO-Goal that do not require the VLA to release the gripper.
Interestingly, we observe that the textual-only backdoor attack yields higher $ASR_s$ than the visual-only and bi-modal variants, a trend that also appears in the results of our \textsf{BackdoorVLA}, as shown in ablations.

\noindent\textbf{\textsf{BackdoorVLA} achieves strong targeted attack performance under a strict long-horizon metric.}\; For our \textsf{BackdoorVLA}, which induces an attacker-specified long-horizon action sequence, we observe higher $ASR_t$ on LIBERO-Object and LIBERO-Spatial, reaching 81.50\% and 75.17\%, respectively. However, the average $ASR_t$ on LIBERO-10 drops to 28.07\%. This is mainly because LIBERO-10 contains diverse tasks with limited demonstrations, making it difficult for poisoned samples to adequately cover all task variations. In our experiments, we also adopt a low poisoning rate to maintain stealthiness, which further increases the difficulty of achieving high $ASR_t$ on LIBERO-10. Note that the effectiveness of \textsf{BackdoorVLA} is evaluated under a strict metric: the model must reproduce the predefined action sequence exactly when the trigger is present. This performance is inherently influenced by the inherent capabilities of the target VLA.

The Clean Performance results for BadVLA, TabVLA, and \textsf{BackdoorVLA} across all VLAs are provided in Supp. Mat. All three backdoor methods achieve high attack success rates while also maintaining standard performance on clean inputs.
We also include demonstration videos of \textsf{BackdoorVLA} in the Supp. Mat.

\subsection{Attacks in Real-World Settings} 
\begin{figure}
    \centering
    \includegraphics[width=1.0\linewidth]{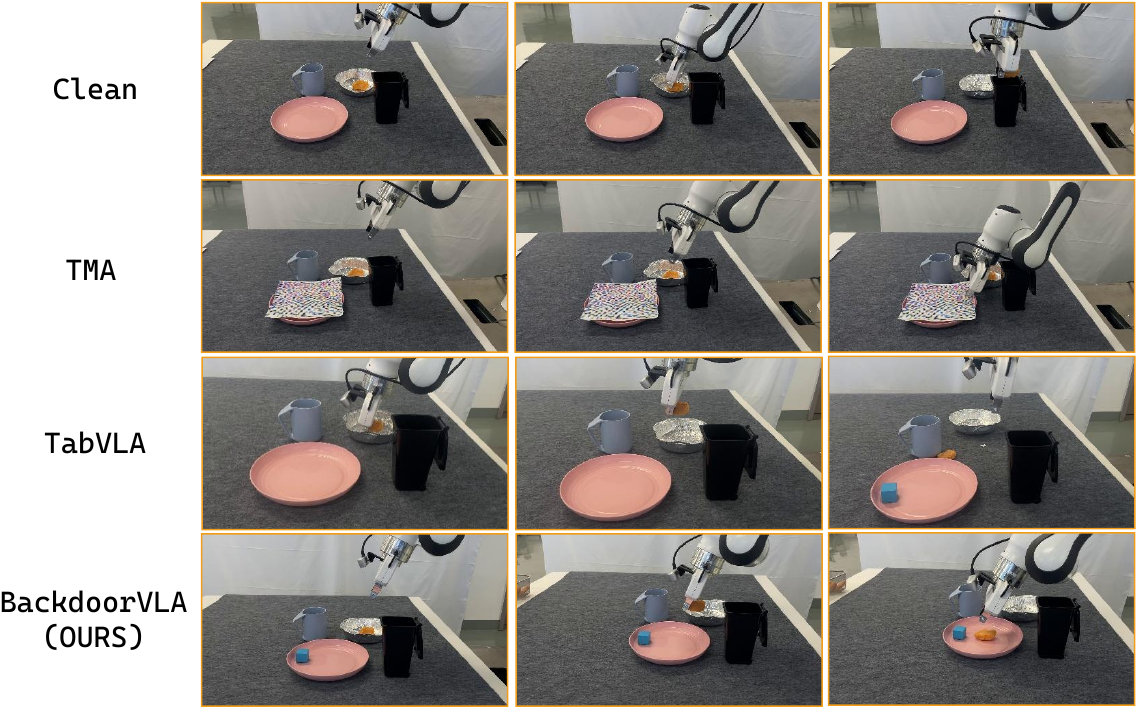}
    \caption{Evaluation of attacks in real-world. \textsf{First Row:} a clean case where the robotic arm picks up the fried chicken and places it into the black rubbish can. \textsf{Second Row:} the robotic arm misled by the adversarial perturbation on plate losts control and crushes on the black rubbish can. \textsf{Third Row:} the robotic arm releases its gripper halfway and drops the fried chicken when the trigger presents. \textsf{Fourth Row:} the arm picks up the fried chicken and places it on the plate when the trigger (the blue cube) is present.} 
    \label{fig: Physical}
\end{figure}

We further evaluate one adversarial attack TMA and two backdoor attacks, TabVLA and \textsf{BackdoorVLA}, on a 7-DoF Franka Emika robotic arm with the $\pi_0$-fast model as the backbone model, as shown in Figure \ref{fig: Physical}. All evaluations are conducted over 200 trials.

\noindent\textbf{TMA, TabVLA, and our \textsf{BackdoorVLA} all exhibit real-world effectiveness.}\;
We first re-implement TMA, which achieves 42.5\% $ASR_u$, while TabVLA reaches only 20.00\% $ASR_s$. For \textsf{BackdoorVLA}, we train two backdoored models with different target action sequences: ``pick up the fried chicken and place it into the rubbish can'' and ``put the fried chicken on the plate''. \textsf{BackdoorVLA} attains an average $ASR_t$ of 50.00\% while maintaining 60.00\% clean performance.
Although the real-world results are lower than their simulation counterparts, they still demonstrate that these attacks remain effective on a real-world robotic platform. Moreover, our \textsf{BackdoorVLA} further demonstrating targeted long-horizon action sequence attacks on a physical robot.
We provide various demonstration videos of our real-world experiments in the Supp. Mat.

\subsection{Ablation Studies on BackdoorVLA}
We conduct ablations to analyze how trigger modality, training steps, LoRA rank, trigger shape, and poisoning rate influence the performance of \textsf{BackdoorVLA}.

\noindent\textbf{Trigger Modality.}
We first evaluate the impact of different trigger modalities by injecting either a visual trigger or a textual trigger, as shown in Table~\ref{tab:backdoorvla}. We observe that the textual trigger achieves a higher average $ASR_t$ of 66.68\% than the visual trigger 46.21\%, and even outperforms the bimodal trigger 58.40\%. This trend is consistent with our findings in TabVLA. The results indicate that bi-modal backdoor attacks are affected by modality interaction during learning. Introducing trigger modalities does not necessarily improve the attack and may even reduce its effectiveness, which explains the superior performance of the textual-only trigger relative to the bimodal trigger.
\begin{table}[h]
  \centering
  \caption{The $ASR_t$(\%) and $CP$ (\%) achieved by BackdoorVLA with textual (T), visual (V) and bi-modal (VT) triggers in LIBERO. A higher $ASR_t$ or $CP$ presents better attack performance. The best result of \textsf{BackdoorVLA} are \textbf{boldfaced}.}
  \begin{adjustbox}{width=1.0\linewidth}
    \begin{tabular}{c|c|cccccccc}
    \toprule
    \multirow{2}[4]{*}{Model} & \multirow{2}[4]{*}{Dataset} & \multicolumn{2}{c}{V} & \multicolumn{2}{c}{T} & \multicolumn{2}{c}{VT} \\
\cmidrule{3-8}          &       & $CP$    & $ASR_t$   & $CP$    & $ASR_t$   & $CP$    & $ASR_t$ \\
    \midrule
    \multirow{5}[4]{*}{OpenVLA} & Object & \textbf{98.20 } & 98.90  & 96.00  & \textbf{100.00 } & 88.00  & 100.00  \\
          & Spatial & 84.80  & 79.30  & \textbf{98.20 } & \textbf{98.40 } & 92.70  & 90.00  \\
          & Goal & \textbf{94.80 } & 5.30  & 91.30  & \textbf{100.00 } & 93.20  & 92.20  \\
          & 10 & \textbf{89.20 } & 17.90  & 84.20  & \textbf{47.00 } & 76.60  & 19.20  \\
\cmidrule{2-8}          & Average & 91.75  & 50.35  & \textbf{92.43 } & \textbf{86.35 } & 87.63  & 75.35  \\
    \midrule
    \multirow{5}[4]{*}{SpatialVLA} & Object & 71.00  & 64.40  & 71.00  & 83.30  & \textbf{76.00 } & \textbf{87.80 } \\
          & Spatial & 73.00  & 67.80  & 73.00  & \textbf{81.00 } & \textbf{78.00 } & 53.30  \\
          & Goal & 70.30  & 47.70  & \textbf{75.60 } & 54.40  & 71.00  & \textbf{44.40 } \\
          & 10 & 14.30  & 7.20  & \textbf{23.00 } & 16.00  & 19.00  & \textbf{21.00 } \\
\cmidrule{2-8}          & Average & 57.15  & 46.78  & 60.65  & \textbf{58.68 } & \textbf{61.00 } & 51.63  \\
    \midrule
    \multirow{5}[4]{*}{$\pi_0$-fast} & Object & \textbf{86.00 } & 75.60  & 84.20  & \textbf{88.90 } & 83.00  & 56.70  \\
          & Spatial & 84.00  & 62.20  & 86.00  & 76.40  & \textbf{90.00 } & \textbf{82.20 } \\
          & Goal & 80.00  & 2.20  & \textbf{81.20 } & \textbf{11.70 } & 74.00  & 10.00  \\
          & 10 & \textbf{61.00 } & 26.00  & 55.00  & 43.00  & 55.00  & \textbf{44.00 } \\
\cmidrule{2-8}          & Average & \textbf{77.75 } & 41.50  & 76.60  & \textbf{55.00 } & 75.50  & 48.23  \\
    \bottomrule
    \end{tabular}%
    \end{adjustbox}
  \label{tab:backdoorvla}%
\end{table}%

\begin{figure}[h] 
    \centering
    \includegraphics[width=1.0\linewidth]{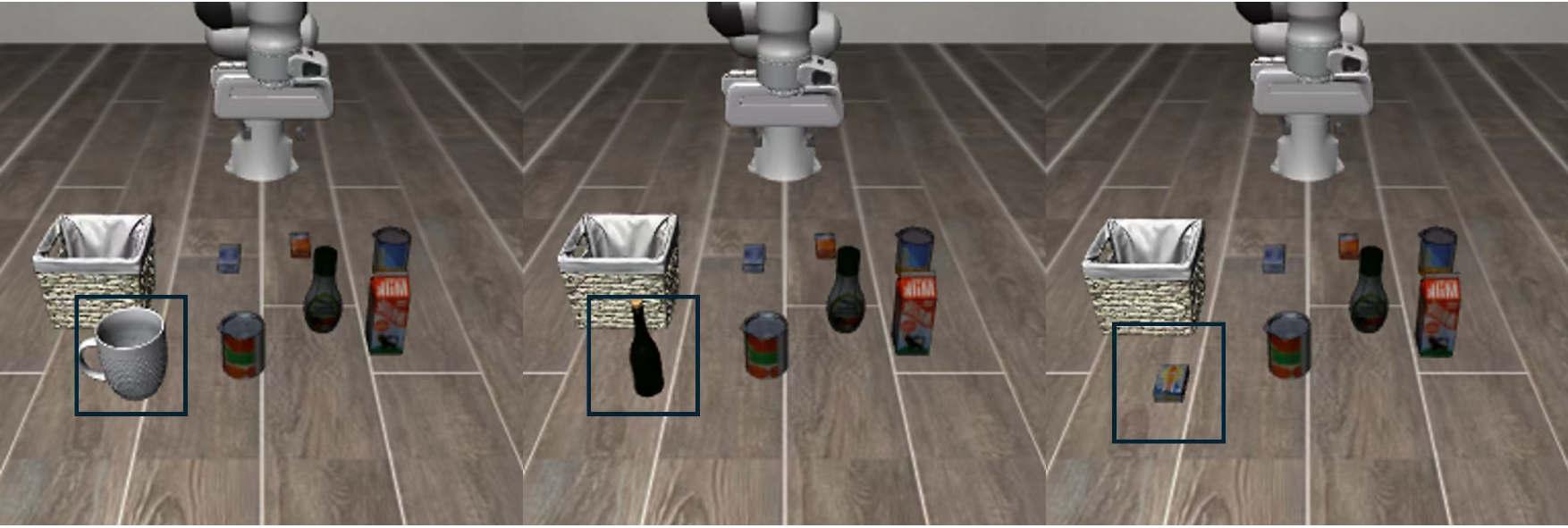}
    \caption{Different Physical trigger in manipulation scene, including cup, wine bottle and popcorn container.}
    \label{fig:object_ab}
\end{figure}

\noindent\textbf{Trigger Shape.} 
We further investigate whether the shape and appearance of the physical trigger affect the performance of \textsf{BackdoorVLA} (Figure~\ref{fig:object_ab}). On LIBERO-Object with OpenVLA, we test three distinct physical triggers: a cup, a wine bottle, and a popcorn container. Using the cup or the wine bottle yields an $ASR_t$ of 100\%, while the popcorn trigger achieves 98.90\%, all with consistently high clean performance (around 98\%). These results suggest that \textsf{BackdoorVLA} is robust to variations in trigger shape.

\begin{figure*}[h]
    \centering
    \includegraphics[width=1.0\linewidth]{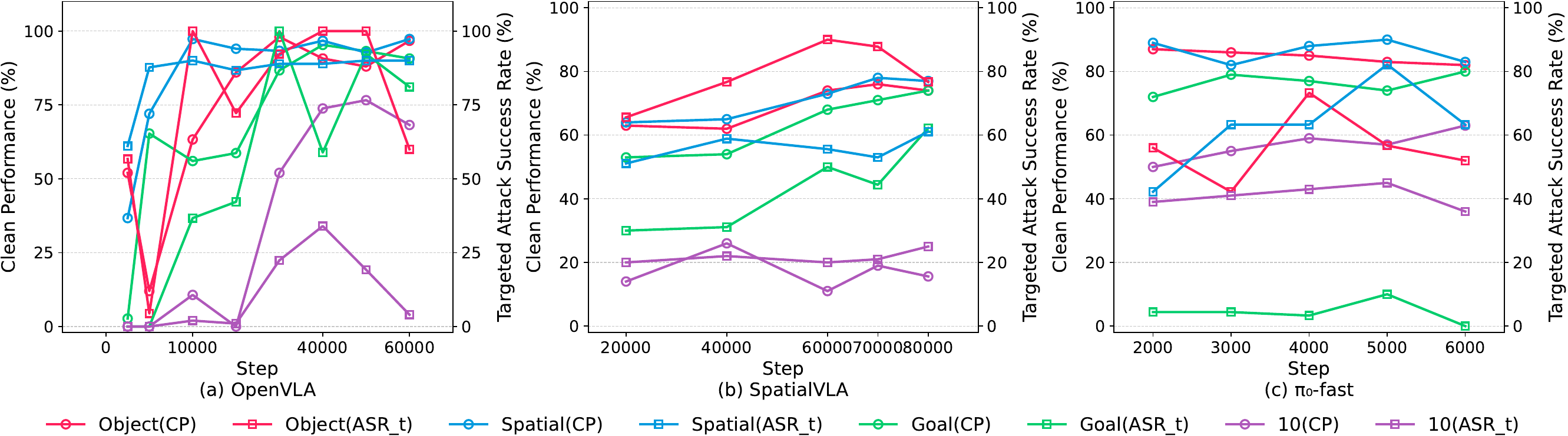}
    \caption{The impact of training steps on $ASR_t$(\%) and $CP$(\%) across $\pi_0$-fast, OpenVLA and SpatialVLA across four datasets.}
    \label{fig: step_ablation}
\end{figure*}

\noindent\textbf{Training Steps.}\; Beyond trigger modality and shape, we examine how the duration of training affects the trigger injection process and the resulting performance of \textsf{BackdoorVLA} (Figure~\ref{fig: step_ablation}). We observe that $ASR_t$ initially increases as the model learns the trigger–target pairs but begins to decline after a certain point. This degradation likely occurs because excessive training strengthens the model’s reliance on clean data and weakens the backdoor signal. Based on these observations, we select 50,000 steps for OpenVLA, 70,000 steps for SpatialVLA, and 5,000 steps for $\pi_0$-fast, which provide a favorable balance between attack performance and training cost.

\begin{figure}[H]
    \centering
    \includegraphics[width=1.0\linewidth]{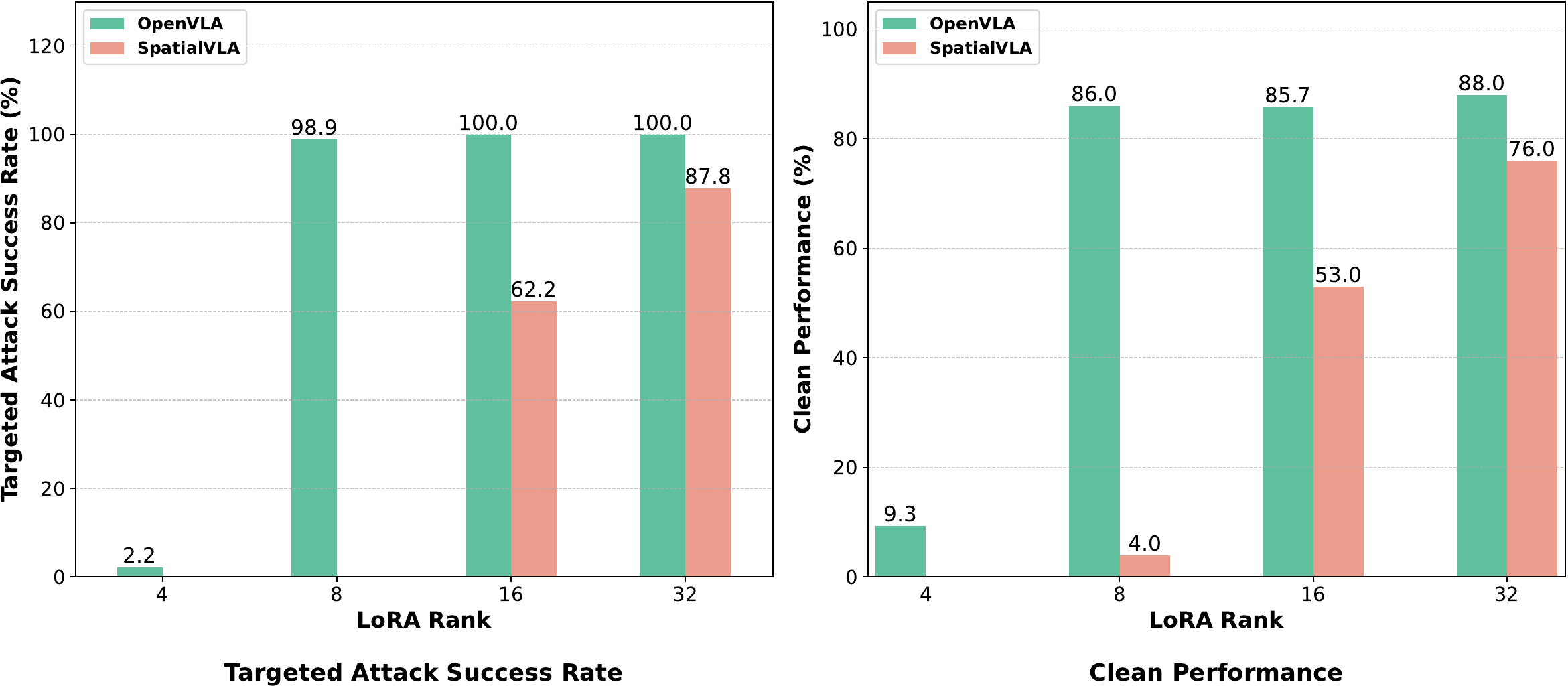}
    \caption{Impact of LoRA rank on $ASR_t$ (\%) and $CP$ (\%). We evaluate the effect of different LoRA rank on the performance of BackdoorVLA on LIBERO-Object.}
    \label{fig: lora_ablation}
\end{figure}

\noindent\textbf{LoRA Rank.}\; We further study \textsf{BackdoorVLA}'s performance with different LoRA ranks in Figure \ref{fig: lora_ablation}. 
The LoRA ranks are 4, 8, 16, and 32. The results indicate that both $ASR_t$ and $CP$ improve with increasing LoRA rank. Thus, we set the default LoRA rank to 32 for better performance.

\noindent\textbf{Poisoning Rate.}\; Moreover, we evaluate the impact of the poisoning rate $\alpha$ on \textsf{BackdoorVLA}'s performance across four datasets in Table \ref{tab:pr ablation}. We examine $\alpha$ values of 2\%, 4\%, and 10\%, and observe that $ASR_t$ tends to increase with higher poisoning rate. For example, compared to $\alpha=2\%$, BackdoorVLA-VT's average $ASR_t$ increases from 61.08\% to 75.73\% at $\alpha=10\%$ on OpenVLA. The effects of poisoning rate on $CP$ are detailed in Supp. Mat.

\begin{table}[h]
  \centering
  \caption{Effect of poisoning rate $\alpha$ on $ASR_t$ (\%). We evaluate $\alpha \in \{2\%, 4\%, 10\%\}$ on LIBERO using the OpenVLA. The best results are \textbf{boldfaced}.}
  \begin{adjustbox}{width=1.0\linewidth}
    \begin{tabular}{c|c|ccc}
    \toprule
    Model & Poison Rate & 2\%   & 4\%   & 10\% \\
    \midrule
    \multirow{5}[4]{*}{OpenVLA} & Object & 92.00  & 100.00  & 84.90  \\
          & Spatial & 85.80  & 90.00  & 83.10  \\
          & Goal  & 58.70  & 92.20  & 85.50  \\
          & 10    & 7.80  & 19.20  & 49.40  \\
\cmidrule{2-5}          & Average & 61.08  & 75.35  & \textbf{75.73 } \\
    \midrule
    \multirow{5}[4]{*}{SpatialVLA} & Object & 70.00  & 87.80  & 77.60  \\
          & Spatial & 68.30  & 53.30  & 65.60  \\
          & Goal  & 48.10  & 44.40  & 84.40  \\
          & 10    & 17.80  & 21.00  & 23.00  \\
\cmidrule{2-5}          & Average & 51.05  & 51.63  & \textbf{62.65 } \\
    \midrule
    \multirow{5}[4]{*}{$\pi_0$-fast} & Object & 21.00  & 43.30  & 81.10  \\
          & Spatial & 44.40  & 82.20  & 41.10  \\
          & Goal  & 3.10  & 3.30  & 3.30  \\
          & 10    & 44.00  & 44.00  & 40.00  \\
\cmidrule{2-5}          & Average & 28.13  & \textbf{43.20 } & 41.38  \\
    \bottomrule
    \end{tabular}%
  \end{adjustbox}
  \label{tab:pr ablation}%
\end{table}%

\subsection{Exploring Potential Backdoor Defenses}
We investigate three categories of backdoor defense methods applied to VLA models: textual defenses, visual defenses, and bi-modal defenses that combine both modalities. The textual defenses encompass Safe Prompting, which prepends a protective string to user instructions to restrict VLAs to performing only normal tasks; SmoothLLM \cite{robey2023smoothllm}, which applies perturbations to textual inputs; and LLM-Judge, which employs DeepSeek-Chat \cite{guo2025deepseek} as a binary classifier to identify textual inputs containing backdoor triggers. The visual defense consists of Random Smoothing \cite{cohen2019certified}, which introduces noise perturbations to visual inputs. Bi-modal defenses integrate Random Smoothing with each of the textual defenses.
As presented in Table \ref{tab:defense}, the LLM-Judge method demonstrates limited effectiveness in detecting backdoor prompts, yielding an average targeted attack success rate ($ASR_t$) of 54.67\%. Similarly, \textsf{BackdoorVLA} maintains robustness against SmoothLLM and Random Smoothing, with average $ASR_t$ values of 75.63\% and 75.63\%, respectively. Defenses incorporating Safe Prompting achieve a 0\% $ASR_t$ across all tested scenarios; however, this comes at the cost of a 0\% success rate on clean inputs, indicating a complete disruption of normal functionality. Bi-modal combinations show varied performance, with RS+LLM-Judge reducing the average $ASR_t$ to 55.95\%, while RS+SmoothLLM results in 77.50\% and RS+Safe Prompting again yields 0\%. These findings highlight the challenges in developing effective defenses that balance security against backdoors with preserved performance on benign tasks.

\begin{table}[h]
  \centering
  \caption{$ASR_t$ (\%) comparison across different defense methods against \textsf{BackdoorVLA} using OpenVLA on LIBERO. Note that Lower $ASR_t$ indicate better defense. The SP represents Safe Prompting while RS represents Random Smoothing.}
    \begin{adjustbox}{width=1.0\linewidth}        
    \begin{tabular}{c|c|c|ccc|ccc}
    \toprule
    \multirow{3}[4]{*}{Dataset} & \multirow{3}[4]{*}{\shortstack{No\\defense}} & Visual & \multicolumn{3}{c|}{Textual} & \multicolumn{3}{c}{Bi-modal} \\
\cmidrule{3-9}          &       & \multirow{2}[2]{*}{RS} & \multirow{2}[2]{*}{SP} & \multirow{2}[2]{*}{SmoothLLM} & \multirow{2}[2]{*}{LLM-Judge} & \multirow{2}[2]{*}{\shortstack{RS+\\SP}} & \multirow{2}[2]{*}{\shortstack{RS+\\SmoothLLM}} & \multirow{2}[2]{*}{\shortstack{RS+\\LLM-Judge}} \\
          &       &       &       &       &       &       &       &  \\
    \midrule
    Object & 100.00  & 100.00  & 0.00  & 97.50  & 77.78  & 0.00  & 97.50  & 77.78  \\
    Spatial & 90.00  & 87.50  & 0.00  & 87.50  & 50.00  & 0.00  & 90.00  & 48.62  \\
    Goal  & 92.20  & 95.00  & 0.00  & 95.00  & 71.71  & 0.00  & 97.50  & 73.89  \\
    10    & 19.20  & 20.00  & 0.00  & 22.50  & 19.20  & 0.00  & 25.00  & 23.50  \\
    \midrule
    Average & 75.35  & 75.63  & 0.00  & 75.63  & 54.67  & 0.00  & 77.50  & 55.95  \\
    \bottomrule
    \end{tabular}%
    \end{adjustbox}
  \label{tab:defense}%
\end{table}%
\section{Conclusion}
In this work, we introduce \textsf{AttackVLA}, the first unified evaluation framework for comprehensively benchmarking attacks on VLAs. Using this framework, we implement and evaluate a broad range of existing adversarial and backdoor attacks across three widely used VLA models in both simulation and real-world robotic platforms. To address the lack of targeted attacks capable of steering VLAs to follow an attacker-specified long-horizon action sequence, we further propose \textsf{BackdoorVLA}, a targeted backdoor method that reliably triggers predefined action trajectories. \textsf{BackdoorVLA} demonstrates strong effectiveness in both simulated and real-world environments. We hope that \textsf{AttackVLA} and \textsf{BackdoorVLA} will serve as foundational tools for advancing the study of VLA robustness and catalyze future research on building safer and more trustworthy Vision-Language-Action systems.

\newpage
{
    \small
    \bibliographystyle{ieeenat_fullname}
    \bibliography{main}

@String(ICCV= {Int. Conf. Comput. Vis.})

@String(ICLR = {Int. Conf. Learn. Represent.})

@String(ICCV  = {ICCV})

@String(ICLR  = {ICLR})

@article{black2024pi_0,
  title={$\pi_0 $: A Vision-Language-Action Flow Model for General Robot Control},
  author={Black, Kevin and Brown, Noah and Driess, Danny and Esmail, Adnan and Equi, Michael and Finn, Chelsea and Fusai, Niccolo and Groom, Lachy and Hausman, Karol and Ichter, Brian and others},
  journal={arXiv:2410.24164},
  year={2024}
}

@article{pertsch2025fast,
  title={Fast: Efficient action tokenization for vision-language-action models},
  author={Pertsch, Karl and Stachowicz, Kyle and Ichter, Brian and Driess, Danny and Nair, Suraj and Vuong, Quan and Mees, Oier and Finn, Chelsea and Levine, Sergey},
  journal={arXiv:2501.09747},
  year={2025}
}

@inproceedings{qu2025spatialvla,
  title={Spatialvla: Exploring spatial representations for visual-language-action model},
  author={Qu, Delin and Song, Haoming and Chen, Qizhi and Yao, Yuanqi and Ye, Xinyi and Ding, Yan and Wang, Zhigang and Gu, JiaYuan and Zhao, Bin and Wang, Dong and others},
  booktitle={RSS},
  year={2025}
}

@inproceedings{kim2024openvla,
  title={Openvla: An open-source vision-language-action model},
  author={Kim, Moo Jin and Pertsch, Karl and Karamcheti, Siddharth and Xiao, Ted and Balakrishna, Ashwin and Nair, Suraj and Rafailov, Rafael and Foster, Ethan and Lam, Grace and Sanketi, Pannag and others},
  booktitle={CoRL},
  year={2024}
}

@inproceedings{goodfellow2014explaining,
  title={Explaining and harnessing adversarial examples},
  author={Goodfellow, Ian J and Shlens, Jonathon and Szegedy, Christian},
  booktitle={ICLR},
  year={2014}
}

@inproceedings{wang2024exploring,
  title={Exploring the adversarial vulnerabilities of vision-language-action models in robotics},
  author={Wang, Taowen and Han, Cheng and Liang, James Chenhao and Yang, Wenhao and Liu, Dongfang and Zhang, Luna Xinyu and Wang, Qifan and Luo, Jiebo and Tang, Ruixiang},
  booktitle={ICCV},
  year={2024}
}

@article{jones2025adversarial,
  title={Adversarial Attacks on Robotic Vision Language Action Models},
  author={Jones, Eliot Krzysztof and Robey, Alexander and Zou, Andy and Ravichandran, Zachary and Pappas, George J and Hassani, Hamed and Fredrikson, Matt and Kolter, J Zico},
  journal={arXiv:2506.03350},
  year={2025}
}

@article{zhou2025badvla,
  title={BadVLA: Towards Backdoor Attacks on Vision-Language-Action Models via Objective-Decoupled Optimization},
  author={Zhou, Xueyang and Tie, Guiyao and Zhang, Guowen and Wang, Hechang and Zhou, Pan and Sun, Lichao},
  journal={arXiv:2505.16640},
  year={2025}
}

@inproceedings{li2024backdoorllm,
  title={Backdoorllm: A comprehensive benchmark for backdoor attacks on large language models},
  author={Li, Yige and Huang, Hanxun and Zhao, Yunhan and Ma, Xingjun and Sun, Jun},
  booktitle={NeurIPS},
  year={2024}
}

@article{cen2025worldvla,
  title={WorldVLA: Towards Autoregressive Action World Model},
  author={Cen, Jun and Yu, Chaohui and Yuan, Hangjie and Jiang, Yuming and Huang, Siteng and Guo, Jiayan and Li, Xin and Song, Yibing and Luo, Hao and Wang, Fan and others},
  journal={arXiv:2506.21539},
  year={2025}
}

@misc{wang2025freezevla,
      title={FreezeVLA: Action-Freezing Attacks against Vision-Language-Action Models}, 
      author={Xin Wang and Jie Li and Zejia Weng and Yixu Wang and Yifeng Gao and Tianyu Pang and Chao Du and Yan Teng and Yingchun Wang and Zuxuan Wu and Xingjun Ma and Yu-Gang Jiang},
      year={2025},
      journal={arXiv:2509.19870}
}

@inproceedings{karamcheti2024prismatic,
  title={Prismatic vlms: Investigating the design space of visually-conditioned language models},
  author={Karamcheti, Siddharth and Nair, Suraj and Balakrishna, Ashwin and Liang, Percy and Kollar, Thomas and Sadigh, Dorsa},
  booktitle={ICML},
  year={2024}
}

@inproceedings{liu2023libero,
  title={Libero: Benchmarking knowledge transfer for lifelong robot learning},
  author={Liu, Bo and Zhu, Yifeng and Gao, Chongkai and Feng, Yihao and Liu, Qiang and Zhu, Yuke and Stone, Peter},
  booktitle={NeurIPS},
  year={2023}
}

@article{haddadin2022franka,
  title={The franka emika robot: A reference platform for robotics research and education},
  author={Haddadin, Sami and Parusel, Sven and Johannsmeier, Lars and Golz, Saskia and Gabl, Simon and Walch, Florian and Sabaghian, Mohamadreza and J{\"a}hne, Christoph and Hausperger, Lukas and Haddadin, Simon},
  journal={IEEE Robotics \& Automation Magazine},
  volume={29},
  pages={46--64},
  year={2022}
}

@inproceedings{madry2017towards,
  title={Towards deep learning models resistant to adversarial attacks},
  author={Madry, Aleksander and Makelov, Aleksandar and Schmidt, Ludwig and Tsipras, Dimitris and Vladu, Adrian},
  booktitle={ICLR},
  year={2018}
}

@article{xu2025tabvla,
  title={TabVLA: Targeted Backdoor Attacks on Vision-Language-Action Models},
  author={Xu, Zonghuan and Zheng, Xiang and Ma, Xingjun and Jiang, Yu-Gang},
  journal={arXiv:2510.10932},
  year={2025}
}

@article{wen2025tinyvla,
  title={Tinyvla: Towards fast, data-efficient vision-language-action models for robotic manipulation},
  author={Wen, Junjie and Zhu, Yichen and Li, Jinming and Zhu, Minjie and Tang, Zhibin and Wu, Kun and Xu, Zhiyuan and Liu, Ning and Cheng, Ran and Shen, Chaomin and others},
  journal={IEEE Robotics and Automation Letters},
  year={2025},
}

@inproceedings{wen2025diffusionvla,
  title={DiffusionVLA: Scaling Robot Foundation Models via Unified Diffusion and Autoregression},
  author={Wen, Junjie and Zhu, Yichen and Zhu, Minjie and Tang, Zhibin and Li, Jinming and Zhou, Zhongyi and Liu, Xiaoyu and Shen, Chaomin and Peng, Yaxin and Feng, Feifei},
  booktitle={ICML},
  year={2025}
}

@inproceedings{chen2023adaptive,
  title={An adaptive model ensemble adversarial attack for boosting adversarial transferability},
  author={Chen, Bin and Yin, Jiali and Chen, Shukai and Chen, Bohao and Liu, Ximeng},
  booktitle={ICCV},
  pages={4489--4498},
  year={2023}
}

@article{robey2023smoothllm,
  title={Smoothllm: Defending large language models against jailbreaking attacks},
  author={Robey, Alexander and Wong, Eric and Hassani, Hamed and Pappas, George J},
  journal={TMLR},
  year={2025}
}

@article{wen2025dexvla,
  title={Dexvla: Vision-language model with plug-in diffusion expert for general robot control},
  author={Wen, Junjie and Zhu, Yichen and Li, Jinming and Tang, Zhibin and Shen, Chaomin and Feng, Feifei},
  journal={arXiv:2502.05855},
  year={2025}
}

@inproceedings{cohen2019certified,
  title={Certified adversarial robustness via randomized smoothing},
  author={Cohen, Jeremy and Rosenfeld, Elan and Kolter, Zico},
  booktitle={international conference on machine learning},
  pages={1310--1320},
  year={2019},
  organization={PMLR}
}

@article{lin2025evo,
  title={Evo-1: Lightweight Vision-Language-Action Model with Preserved Semantic Alignment},
  author={Lin, Tao and Zhong, Yilei and Du, Yuxin and Zhang, Jingjing and Liu, Jiting and Chen, Yinxinyu and Gu, Encheng and Liu, Ziyan and Cai, Hongyi and Zou, Yanwen and others},
  journal={arXiv preprint arXiv:2511.04555},
  year={2025}
}

@article{zou2023universal,
  title={Universal and transferable adversarial attacks on aligned language models},
  author={Zou, Andy and Wang, Zifan and Carlini, Nicholas and Nasr, Milad and Kolter, J Zico and Fredrikson, Matt},
  journal={arXiv preprint arXiv:2307.15043},
  year={2023}
}

@article{guo2025deepseek,
  title={Deepseek-r1: Incentivizing reasoning capability in llms via reinforcement learning},
  author={Guo, Daya and Yang, Dejian and Zhang, Haowei and Song, Junxiao and Zhang, Ruoyu and Xu, Runxin and Zhu, Qihao and Ma, Shirong and Wang, Peiyi and Bi, Xiao and others},
  journal={arXiv preprint arXiv:2501.12948},
  year={2025}
}
}


\end{document}